\documentclass[pageno,nohyperref]{jpaper}

\usepackage{amsmath}
\usepackage{todonotes}
\usepackage{graphicx}
\usepackage{booktabs}
\usepackage{array}
\usepackage{authblk}
\usepackage{enumitem}

\newcommand*{\ARXIV}{}

\unless\ifdefined\ARXIV
    \graphicspath{{./images/}}
\fi

\title{Optimizing the flash-RAM energy trade-off in deeply embedded systems}

\author[1]{James Pallister\thanks{james.pallister@bristol.ac.uk}}
\author[1]{Kerstin Eder\thanks{kerstin.eder@bristol.ac.uk}}
\author[1]{Simon J. Hollis\thanks{simon.hollis@bristol.ac.uk}}
\affil[1]{University of Bristol}
\date{}

\begin{document}

\maketitle

\begin{abstract}
    Deeply embedded systems often have the tightest constraints on energy consumption, requiring that they consume tiny amounts of current and run on batteries for years. However, they typically execute code directly from flash, instead of the more energy efficient RAM. We implement a novel compiler optimization that exploits the relative efficiency of RAM by statically moving carefully selected basic blocks from flash to RAM. Our technique uses integer linear programming, with an energy cost model to select a good set of basic blocks to place into RAM, without impacting stack or data storage.

    We evaluate our optimization on a common ARM microcontroller and succeed in reducing the average power consumption by up to 41\% and reducing energy consumption by up to 22\%, while increasing execution time. A case study is presented, where an application executes code then sleeps for a period of time. For this example we show that our optimization could allow the application to run on battery for up to 32\% longer. We also show that for this scenario the total application energy can be reduced, even if the optimization increases the execution time of the code.
\end{abstract}

\section{Introduction}

Deeply embedded System on Chips (SoCs) are prevalent in our portable devices. These SoCs are small microcontrollers, without caches that typically contain RAM and flash. Code is executed directly from the flash and the RAM used for volatile data storage (initialized on startup).

Many of these devices are low speed, allowing both flash and RAM to be accessed in a single cycle. However, these memories have very different energy consumption characteristics. Reading from flash memory typically takes more power than reading from RAM. This means that if code can be executed from RAM instead of flash, there should be a marked improvement in energy consumption.

\begin{figure}[b!]
    \centering
    \includegraphics[width=.9\linewidth]{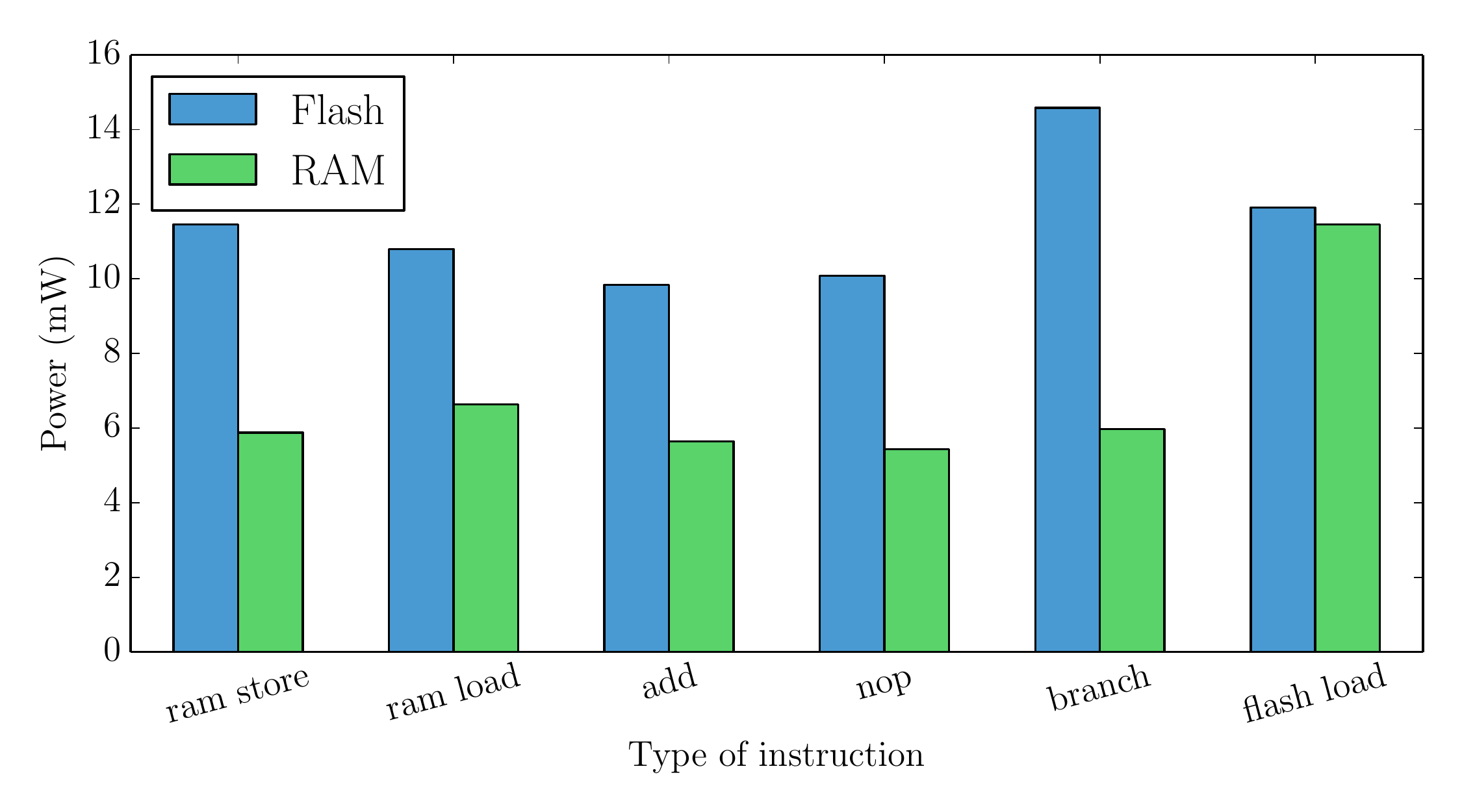}
    \caption{Average power for different instructions, when executing out of flash and RAM.}
    \label{fig:comparison}
\end{figure}

\begin{figure*}
    \centering
    \includegraphics[width=.99\linewidth]{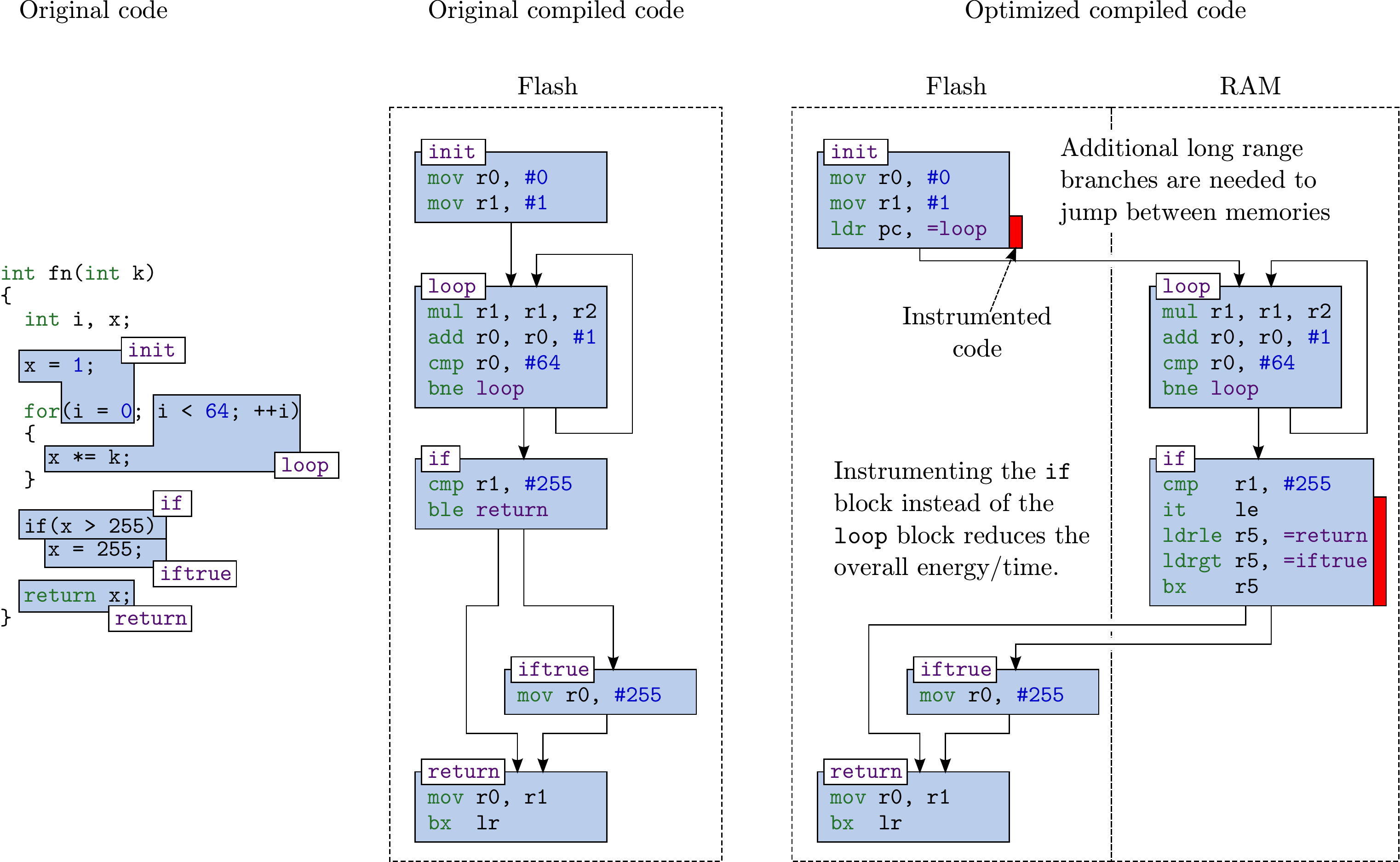}
    \caption{An example of a function (left), its original compiled version (center), and the optimized version with basic blocks in RAM (right). The given code is for the Cortex-M3.}
    \label{fig:examplecfg}
\end{figure*}

Figure~\ref{fig:comparison} compares a set of extremely simple programs run in flash and RAM. These programs consist of 16 identical instructions in a loop. This loop is placed in flash and then in RAM, showing the difference in power consumption. The power consumption is significantly lower when the code is executing from RAM, except when the code in RAM also accesses the flash (for example to load constant data), as seen in the last bar of Figure~\ref{fig:comparison}. This motivates moving as much code as possible into RAM.

Unfortunately, it is not economical to have equal amounts of flash and RAM in deeply embedded systems --- typically the there is an 8:1 (or higher ratio) of flash to RAM. Therefore, not all of the code can be placed into RAM especially as the RAM must also contain the volatile data and stack. Only the most effective parts of code can be moved, to minimize the energy consumption while still being constrained to the available amount of RAM.

A motivating example is given in Figure~\ref{fig:examplecfg}. This example shows a function and its control flow graph, for which the inner loop will be executed much more frequently than its surrounding basic blocks (a basic block is a sequence of code in which the control flow enters or exits only from the beginning or end, respectively). It may not be possible to move the entire function into RAM. However, it may be possible to move just some basic blocks. Because the loop is executed more frequently, the relative gain from moving it is larger. The block following the loop is also moved into the RAM, as this negates the need to add the long range branch to the inner loop.

This problem has been tackled in a similar form by the scratchpad memory community. Many of these techniques are transferable, and a review of these techniques is given in Section~\ref{sec:related_work}. In this paper, Integer Linear Programming (ILP)~\cite{Hoffman2013} is used with a basic block cost model to find a set of basic blocks which would benefit from being in RAM. Individual basic blocks are statically placed into RAM, rather than full functions. This allows better use to be made of the limited RAM in these deeply embedded systems, by only placing energy intensive basic blocks into RAM. The optimization is applicable to any microcontroller with a unified address space --- the ability to transfer control flow into RAM is necessary to implement the optimization.


In this paper, this is implemented on a power-measurement instrumented Cortex-M3 processor~\cite{STM32F100}. This processor is a commonly used microcontroller. It has a unified address space, allowing instructions to be executed from either flash or RAM. The SoC has 64KB of flash and 8KB of RAM. This SoC is frequently used in low power applications and these applications could directly benefit from the techniques introduced in this paper.

\noindent
This paper makes the following contributions to the areas of embedded energy efficiency and compiler optimization:
\begin{itemize}
    \item The design of a novel optimization which analyzes the program to find a set of basic blocks which should be transferred into RAM to improve the energy consumption. This involves the construction of a model describing the costs associated with moving this set of basic blocks into RAM. The optimization rewrites the branches at the end of basic blocks that need to jump between RAM and flash.
    \item The evaluation of this optimization on BEEBS~\cite{Pallister2013b}, a benchmark suite designed to allow testing of embedded systems for energy consumption. The model is evaluated by examining the solutions it selects, compared to a large sample of possible solutions.
    \item A realistic case example of periodic sensing is presented, where a device wakes from sleep to perform computation, then returns to sleep. This presents the unintuitive result that when the energy of the active region is not reduced and the execution time is increased, then the overall energy consumption of the application can still be decreased.

    In this scenario, the lower energy consumption and higher execution time actually benefit the application, calculating that the device's battery life can be extended by up to 32\%.
\end{itemize}

In the following section, related work is discussed. Then, the optimization's general methodology is presented. Then, the methodology is discussed in more detail, firstly with the cost model (Section~\ref{sec:model}) and then the code transformation (Section~\ref{sec:code_transformation}). Section~\ref{sec:evaluation} describes the tests and efficacy of the optimization. Then, Section~\ref{sec:case_study} presents a case study which examines the optimizations effectiveness in a real world situation. Finally, the paper is concluded in Section~\ref{sec:conclusion}.

\section{Related work}
\label{sec:related_work}

The problem of moving parts of code and data from one memory to a faster memory has been studied extensively in the context of scratchpad memory. Most studies focus on static assignment of code and data to the scratchpad memory with the aim of decreasing program execution time or energy consumption. Steinke et al.~\cite{Steinke2002} compare scratchpad memories and caches, finding that a scratchpad memory can save up to 43\% of the energy consumption compared to a cache of the same size. This is achieved by using Integer Linear Programming (ILP) to minimize the cost of placing basic blocks into memory. There have been many different formulations of ILP problems, considering data objects~\cite{Sjodin1998}, individual basic blocks~\cite{Wehmeyer2004} and cache-awareness~\cite{Verma2006}. Ishitobi et al.~\cite{Ishitobi2008} use ILP in a system with both caches and scratchpad memory, creating a model to decide whether the particular item should be placed in either a cacheable or scratchpad memory region. This reduces both energy consumption and execution time.


Other work on scratchpad memory has attempted to dynamically move objects into memory as they are needed~\cite{Steinke2002a}. This study identified which parts of the code should remain in scratchpad memory and which parts should be brought in dynamically at specific locations through the program. Another study~\cite{Verma2004} applies techniques developed for global register allocation to scratchpad memories, reducing the energy consumption by up to 34\%. In both of these studies, the energy saving is partly due to the decrease in execution time.

A different approach is taken by Kandemir et al.~\cite{Kandemir2001a}, where Presburger formulae are used to minimize the number of transfers between main memory and the scratchpad memory. This technique manages to reduce the number of off-chip references and memory energy consumption.

Sharing a scratchpad memory between multiple tasks has been tackled in~\cite{Gauthier2010}, by attempting to optimally pack different task's regions of code and data into the scratchpad memory.

Many other scratchpad memory allocation schemes have been proposed. A comprehensive review of these is given in~\cite{ScratchpadMemoryChapter2006}, including multiple scratchpad memories, and partitioned memories.

For embedded systems with no scratchpad memory, but just flash and RAM, there has been less extensive research. Park et al.~\cite{Park2006} attempt to minimize the amount of RAM required for programs executing directly from NAND flash by using a dynamic code overlay to execute the code from RAM. Execution out of flash has been optimized using a page manager to copy pages of flash into RAM at runtime, with analysis support provided by the compiler~\cite{Park2004a}. Our study instruments branches in a similar way, however in deeply embedded systems both memories are single cycle access, so the dynamic approach is not needed.


Software-level energy modeling was first discussed by Tiwari et al.~\cite{Tiwari1996}. This model consisted of an energy cost for each instruction, an energy cost for each transition between sequential instructions and a term for extra effects (such as caches). This allowed simulation traces to estimate the energy consumption without requiring to be run on physical hardware. The method was refined into a fine grained model also accounting for the energy due to differing data in~\cite{Steinke2001}. These approaches have been further developed to allow architectural exploration, with Wattch~\cite{Brooks2000} and SimplePower~\cite{Ye2000}. More recent studies have explored how these models can be extended to processors with hardware multithreading~\cite{Kerrison2013}.

Modeling energy consumption has been explored at the function level of code~\cite{Qu2000}. This involves creating a `data bank' of how much energy each function costs to run. These energy figures can then be distributed with libraries, or combined with instruction level modeling~\cite{Blume2006} to estimate a programs energy consumption.

Energy modeling has also been explored at a higher level, by considering the average power of each state the processor can be in~\cite{Min2012,NunezYanez2013}. This requires less knowledge about the exact instruction stream of the processor, simply the times spent in each mode.


\section{Methodology}

Much of the previous work has focused on devices that have a scratchpad memory operating several times quicker than the data or instruction memory. The systems targeted by this work do not have a scratchpad memory, just a unified address space with embedded flash and RAM. Both memories are single cycle access, resulting in no performance gain if the code is moved out of flash and into RAM. These memory access times causes a net performance loss when code is distributed across both flash and RAM, however, should save energy due to RAM's lower average power.

A cost model is created to describe the effect of moving different regions of code from flash to RAM. This model considers the energy consumption in a simplified way --- an average power is assigned to each instruction, based on whether the instruction executes out of RAM or flash. The cycle cost of each basic block is modeled in more detail, since moving code from flash to RAM results in the code taking additional execution cycles. The overhead is integrated into the model, enabling the developer to set a maximum slow down. This allows multi-objective optimization, balancing the trade-off between code size, energy and performance.

The total energy sum is minimized by an ILP solver. This solver uses the cost model and parameters extracted from the program to estimate which are the optimal basic blocks of the program to place into memory.

The set of basic blocks to be placed in RAM is used to transform the code, moving the correct basic blocks to a section which is loaded to RAM. The basic blocks which jump between memories are instrumented such that they have the range to make the long jump.

The resultant code is run on physical processors. The majority of studies rely on models, which may not incorporate all effects the hardware may have on the energy. Thus actual measurements are taken to verify the efficacy of the final code. Effects such as position dependent energy consumption in memory~\cite{Pallister2014} and large variability between supposedly identical processors~\cite{Wanner2010} necessitate evaluation with real, instrumented hardware.

\section{Formulating the ILP model}
\label{sec:model}

In this section a model and set of constraints is developed to allow optimal selection of which parts of code are moved into RAM. This model is heavily influenced by~\cite{Steinke2002}, extending and improving it in the following ways:

\begin{itemize}
    \item The cost of modifying the code in the basic block to branch between memories is accurately described. This results in the solver automatically `clustering' small basic blocks into RAM, which would otherwise have caused a large overhead when branching between memories.
    \item The model is based on the number of cycles spend executing from RAM and flash as the cost metric, instead of the number of instructions. This is necessary since the Cortex-M3 attempts to prefetch instructions as well as speculatively fetching branch destinations~\cite{Yiu2010}. Having the cycle count of the code accounted for means that the overhead in execution time will also be minimized.
    \item There is no need to consider data items in the model, since the volatile data items are already in RAM (copied on startup by the runtime). Constant data is still stored in the flash, however this is typically accessed infrequently, and mostly used for initialization.
\end{itemize}

There are a couple of factors which affect the efficacy of placing a basic block in RAM.
One factor is the overhead of instrumenting the basic block to be able to jump to the other memory. Each transition between flash and RAM must be done with an indirect branch, rather than a typical direct branch. Indirect branches allow the long ranging jumps between memory spaces to be made. However they typically take longer to execute, or require other supporting instructions. The instrumentation overhead is discussed further in Section~\ref{sec:code_transformation}. This instrumentation is only performed if one of the basic block's successors is in a different memory space --- the instrumentation is not needed if the block's successors are in the same memory.

The relative benefit of placing the block into RAM must also be considered. It is more beneficial to place frequently executed blocks in RAM, since the reduction in energy consumption will be greater. However, it is also beneficial to place blocks in RAM if it removes the need to instrument a frequently executed block. It is for this reason that small joining basic blocks between frequently executed loops may be moved into RAM.

The rest of this section is divided into a discussion of the parameters required by the model, and then how the model is constructed from these parameters.

\subsection{Parameters}

\begin{figure}
    \centering
    \includegraphics[width=0.9\linewidth]{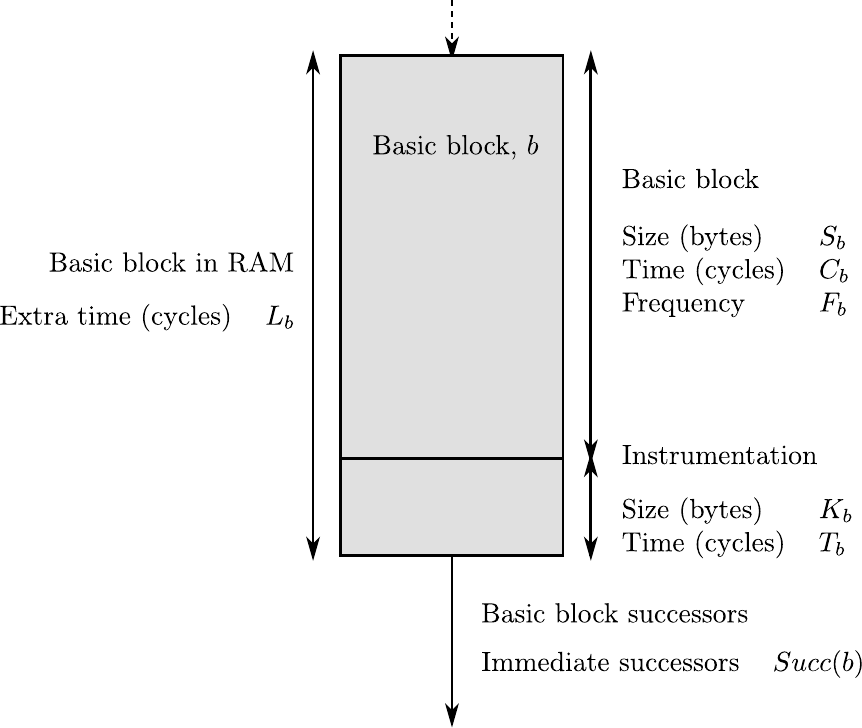}
    \caption{Model parameters for a basic block, $b$.}
    \label{fig:parameters}
\end{figure}

Several parameters are given to the model, allowing a solver to consider most of the factors which will affect the energy consumption of the code. This section discusses the parameters required by the model, and how they are calculated.

The following parameters are derived automatically from the structure of the source code. The parameters differ for each basic block. A diagram showing these parameters is given in Figure~\ref{fig:parameters}.

\vspace{1mm}

\begin{description}[labelindent=2mm, labelsep=5mm, listparindent=0mm, itemindent=-4mm, parsep=1mm]
    \item[$S_b$]
        This parameter describes the size of the basic block, $b$, in bytes.
    \item[$C_b$]
        The number of cycles taken to execute the basic block. This will always be a best estimate, due to complexities in the processor, such as fetch stalling and pipeline flushes when conditional branches are taken.
    \item[$F_b$]
        This is the `frequency' of the basic block --- the number of times it is executed. This can be found by profiling the application, or by statically analyzing the code.

        A simple estimate can be made of this parameter by simply considering the block's loop-depth. Section~\ref{sec:evaluation} discusses how estimates of this parameters affect the final solution, showing that a rough estimate is good enough in most cases.
    \item[$K_b$]
        The instrumentation cost of the basic block, in bytes. This is the number of necessary extra bytes to instrument the basic block with jumps between memory spaces.
    \item[$T_b$]
        The instrumentation cost of the basic block, in cycles. This is the number of additional cycles executed when the basic block is instrumented with jumps between memory spaces.
    \item[$L_b$]
        This is the number of additional cycles required when the block is in RAM. This stems from contention on the memory bus, and is proportional to the number of load instructions in the basic block.
    \item[$Succ(b)$]
        This specifies the set of basic blocks which are immediate successors to the block, $b$.
\end{description}

\vspace{3mm}
\noindent
The following parameters are specified by the developer.
\vspace{1mm}
\begin{description}[labelindent=2mm, labelsep=5mm, listparindent=0mm, itemindent=-4mm, parsep=1mm]
    \item[$X_{limit}$]
        This is a `time factor', indicating the maximum overhead that should be allowed in the solution. For example, setting $X_{limit}=1.1$ allows the solver to pick a combination of blocks to go in RAM that should take less than 10\% longer to execute.
    \item[$R_{spare}$]
        The maximum amount of RAM to use for code. The solver can be restricted to using fewer bytes of RAM, to fit within memory limits. This can also be derived statically, by considering the size of the variables in RAM, heap and the stack usage~\cite{Brylow2001}.
\end{description}

\vspace{3mm}
\noindent
The following parameters are determined from the hardware.

\vspace{1mm}
\begin{description}[labelindent=2mm, labelsep=4mm, listparindent=0mm, itemindent=-8.5mm, parsep=1mm]
    \item[$E_{flash}$]
        A coefficient representing the energy cost of executing out of flash. The average power when executing instructions out of flash is assigned to this parameter.
    \item[$E_{ram}$]
        A coefficient representing the energy cost of executing out of RAM. The average power when executing instructions out of RAM is assigned to this parameter.
\end{description}

\subsection{Auxiliary parameters/functions}

\noindent
The following sets are determined during the solving.
\vspace{1mm}
\begin{description}[labelindent=2mm, labelsep=4mm, listparindent=0mm, itemindent=-2mm, parsep=1mm]
    \item[$\bf B$]
        The set of all basic blocks. This is extracted from the control flow graph.
    \item[$\bf R$]
        This set is the set of basic blocks that are moved into RAM.
    \item[$\bf I$]
        This set is the set of basic blocks that need to be instrumented, since one of their successors is present in a different memory. This is purely for convenience, and can be calculated purely from $\bf R$.
\end{description}

\vspace{3mm}
\noindent
Several other functions are based on the given parameters.

\vspace{1mm}
\begin{description}[labelindent=2mm, labelsep=4mm, listparindent=0mm, itemindent=-8.5mm, parsep=1mm]
    \item[$M(b)$]
        This function returns memory energy cost, depending on whether $b$ is in RAM or not.
    \item[$O_c(b)$]
        A function that returns the cycle overhead for the basic block, $b$. If the block is not instrumented then this is $0$.
    \item[$O_r(b)$]
        This function returns the cycle overhead if the basic block is in RAM. This occurs because of contention on the memory bus when a load instruction is encountered.
    \item[$O_s(b)$]
        A function which returns the space overhead for the basic block, $b$, when the block is instrumented. If the block is not instrumented then this is $0$ for this particular basic block.
\end{description}

\subsection{The model}

The problem is formulated as a minimization of the total energy of the program, by finding a set of basic blocks, $\bf R$, which represents the set of basic blocks placed into RAM.

\begin{equation}
    \mathrm{Minimize:\quad}{} \sum_{b\in {\bf B}}E(b),
\end{equation}
where $\bf B$ is the set of all basic blocks. The energy of each basic blocks can be determined,

\begin{equation}
    E(b) = \big(C_b + O_c(b) + O_r(b)\big)\cdot M(b) \cdot F_b,
\end{equation}
where $C_b$ is the number of cycles the basic block takes to execute, $O_c(b)$ is the cycle overhead added to the basic block (dependent on whether $b$ is placed into RAM or not), $M(b)$ is an energy coefficient, describing the energy cost per cycle of executing out of a particular type of memory. $F_b$ is an execution frequency for that particular block, scaling the energy cost of $b$ by how many times it is executed. $O_r(b)$ is the cycle overhead from a block being in RAM.

The cycle count, $C_b$, and the execution frequency, $F_b$, of the block are input variables into the equation. In previous works, the number of instructions is used in place of the cycles per basic block~\cite{Steinke2002}. However, under most circumstances this processor (the Cortex-M3) will fetch every cycle, filling the prefetch buffer during a multi-cycle instruction.

The memory energy cost of a basic block is easily determined --- it is only dependent on whether the basic block is in RAM or not:

\begin{equation}
    M(b) = \left\{
   \begin{array}{ll}
        E_{ram}     & b\in \bf R  \\
        E_{flash}   & b\notin \bf R, \\
    \end{array} \right.
\end{equation}
where $\bf R$ is the set of basic blocks which are in RAM, and $E_{ram}$ and $E_{flash}$ are coefficients describing the energy cost of RAM and flash respectively.

If the block needs to be instrumented, then the instrumentation overhead needs to be factored in,

\begin{equation}
    O_c(b) = \left\{
    \begin{array}{ll}
        T_b     & b \in \bf I  \\
        0       & b \notin \bf I, \\
    \end{array} \right.
\end{equation}
where $T_b$ is an instrumentation overhead for that block, in cycles, and $\bf I$ is the set of basic blocks which need to be instrumented. A basic block needs to be instrumented if it and any of its successors are not in the same memory space.

\begin{equation}
    \begin{array}{l}
    b \notin {\bf I} \mathrm{\quad if\quad} b \in {\bf R} \textrm{\hspace{1.5mm}and\hspace{1.5mm}} \forall \big(x \in Succ(b)\big): x\in {\bf R} \\
    b \notin {\bf I} \mathrm{\quad if\quad} b \notin {\bf R} \textrm{\hspace{1.5mm}and\hspace{1.5mm}} \forall \big(x \in Succ(b)\big): x\notin {\bf R} \\
    b \in {\bf I} \mathrm{\quad otherwise,} \\
    \end{array}
\end{equation}
where $Succ(b)$ returns a set of all blocks which are immediate successors to $b$, and can be extracted from the control flow graph.

Although both the flash and the RAM are nominally single cycle access, there are cases when the number of cycles can differ. The most prevalent case is when executing a load from RAM, while executing out of RAM. The processor stalls for extra cycles in this case, due to contention for the RAM memory interface. This contention does not occur when executing out of flash, due to there being a separate memory interface for flash. The cycle overhead when is executing from RAM is given by:

\begin{equation}
    O_r(b) = \left\{
    \begin{array}{ll}
        L_b     & b \in {\bf R}  \\
        0       & b \notin {\bf R}, \\
    \end{array} \right.
\end{equation}
where $L_b$ is the number of stall cycles due to load instructions being executed when in RAM ($b\in {\bf R}$).

These definitions form the basis of the model, however additional constraints are added to ensure that the maximum amount of spare RAM is not exceeded, and that the execution time does not grow larger than optimal. The constraint on RAM usage is given,

\begin{equation}
    \sum_{b\in {\bf R}}(S_b + O_s(b)) \leq R_{spare},
\end{equation}
where $R_{spare}$ is the amount of `spare' RAM which the program does not use and $O_s(b)$ is the amount of RAM necessary to instrument a block (bytes). In many cases for embedded systems, the spare RAM can be determined statically, by stack analysis. Otherwise, the RAM dedicated to program code will be a design decision, similar to the amount of stack and heap space to allocate to the program.

The size instrumentation cost of the basic block is similar to $O_c(b)$:

\begin{equation}
    O_s(b) = \left\{
    \begin{array}{ll}
        K_b     & b \in {\bf I}  \\
        0       & b \notin {\bf I}, \\
    \end{array} \right.
\end{equation}
where $K_b$ is the instrumentation cost of $b$ in bytes.

\begin{figure*}
    \includegraphics[width=\textwidth]{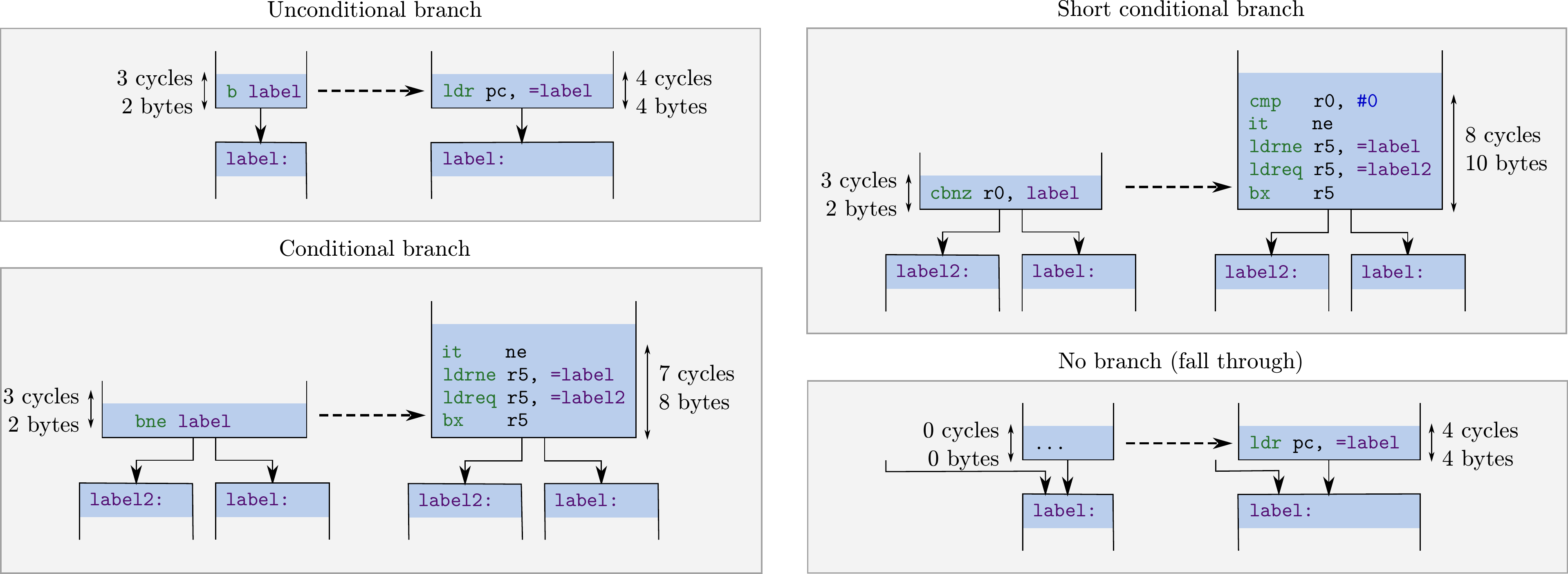}
    \caption{The transformations applied to the basic blocks which need to be instrumented ($b\in {\bf I}$). The text gives the execution time (cycles) and the size (bytes) of the code sequences.}
    \label{fig:transformations}
\end{figure*}

The execution time constraint can be formulated in a similar way by considering the number of cycles that each basic block requires to execute. The ratio of original execution time to execution time with overhead can be constrained, ensuring that the execution time does not grow by more than $X_{limit}$:

\begin{equation}
        \frac{\sum_{b\in {\bf B}}\big((C_b + O_c(b) + O_r(b))\cdot F_b\big)}{\sum_{b\in {\bf B}}(C_b\cdot F_b)} \leq X_{limit},
\end{equation}
where $C_b$ is the number of cycles the block $b$ takes to execute, $O_c(b)$ is the cycle overhead from block instrumentation, and $O_r(b)$ is the cycle overhead from a block being in RAM.




This model can be linearized so that it can be solved with a standard ILP solver. In this paper the GNU Linear Programming Kit (GLPK)~\cite{GLPK} is integrated into the optimization. This solver returns the set of basic blocks that should be in RAM.

\section{Code transformation}
\label{sec:code_transformation}

Once the basic blocks to be in RAM are chosen, the transformation can relocate these blocks and instrument the necessary blocks. The actual transformation itself happens at the very end of compilation. The parameters for the model are extracted from the CFG, passed to the solver and the basic blocks are modified in accordance to whether they are in RAM or not. All basic blocks which are required to be in RAM are moved into a custom section of the executable which is loaded into RAM at start-up by the runtime.

The transformation also modifies the basic blocks, based on whether they are required to jump between memory spaces or not. In general, there are three basic forms the instrumentation takes, each corresponding to the type of jump at the end of a basic block:

\begin{description}
    \item[Unconditional branch.] If the branch at the end of the block is unconditional this just needs to be exchanged to an indirect branch, loading the target address from memory. This enables a much larger branch range, necessary for the jump into RAM or back into flash.
    \item[Conditional branch.] A basic block with a conditional branch at the end could transfer execution to two possible locations. Both must be instrumented. If the branch is not taken, the execution falls through into the following block. Here, an indirect branch must be added, since the following block may not be in the same memory space.
    \item[No branch.] As with a conditional branch, if the following basic block is not in the same memory space the transition must be instrumented. This is simply an indirect branch.
\end{description}

The specific code changes made to instrument a basic block for the Cortex-M3 are shown in Figure~\ref{fig:transformations}. This illustrates how the code is modified for each type of basic block, along with the space and time overheads for doing so. For the Thumb2 instruction set~\cite{ARMLimited2010a} used by the Cortex-M3 there is an additional type of conditional branch which requires slightly different instrumentation, due to it combining the comparison into the instruction.

\section{Evaluation}
\label{sec:evaluation}

\begin{figure}
    \includegraphics[width=\linewidth]{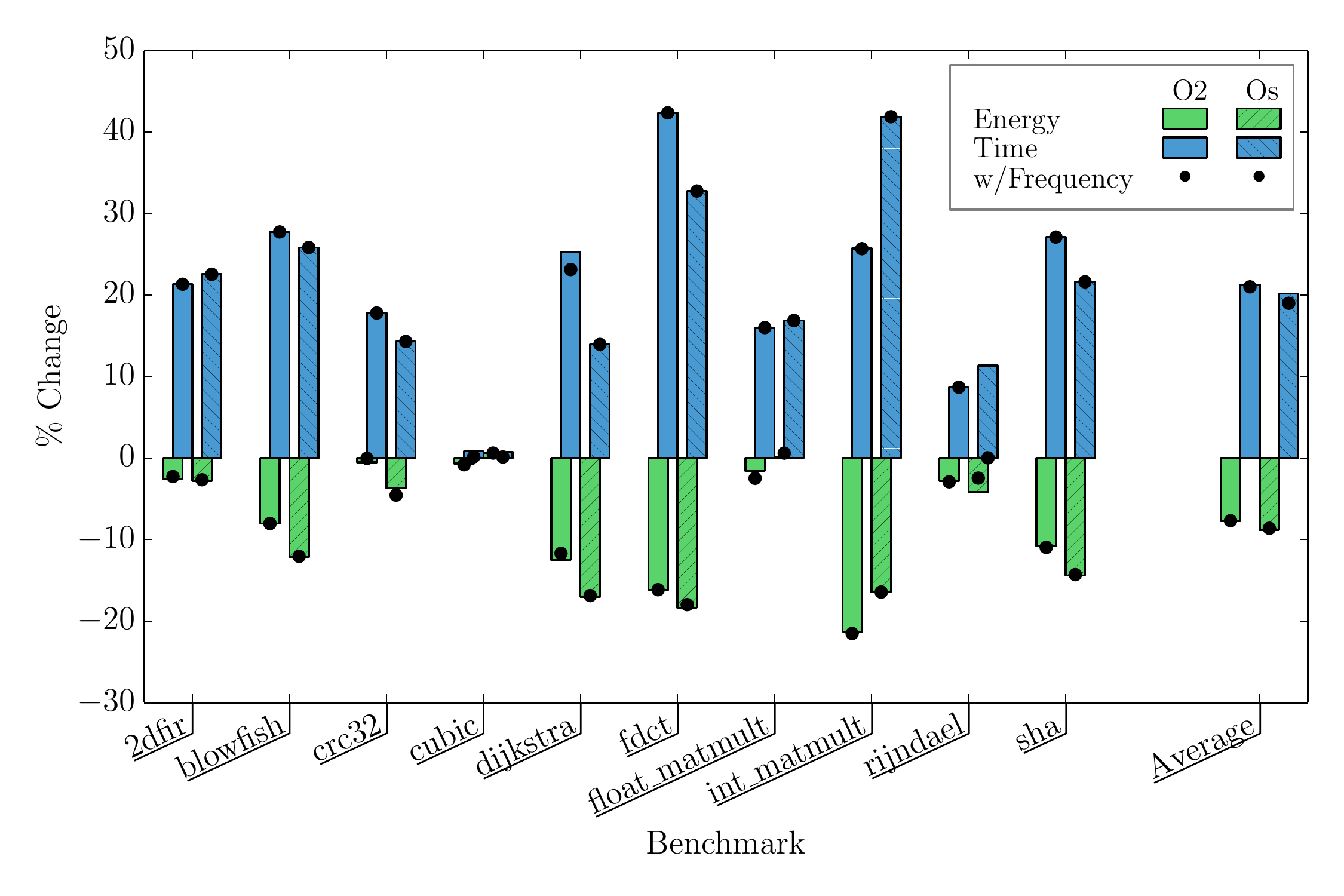}
    \caption{Results for applying the optimization pass on the BEEBS benchmark suite at different optimization levels. Each pair of energy and time bars is a single run of the benchmark. The dots indicate an additional run with actual basic block frequencies as opposed to the estimate.}
    \label{fig:results}
\end{figure}

\begin{figure*}[t]
    \subfloat[\texttt{int\_matmult}\label{fig:exhaustive_int_matmult}]{
        \includegraphics[width=0.49\linewidth]{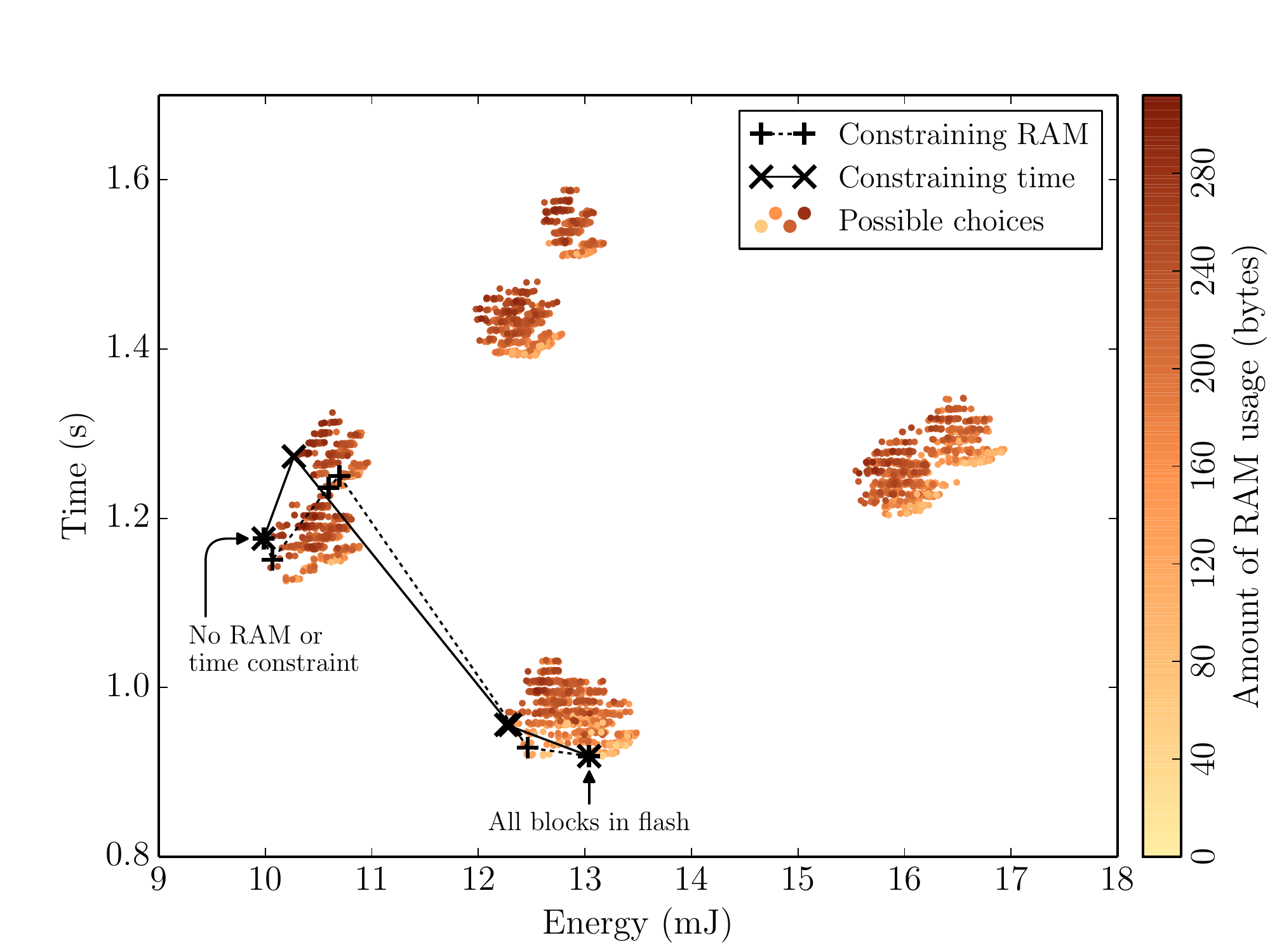}%
        }
    \hfill
    \subfloat[\texttt{fdct}\label{fig:exhaustive_fdct}]{
        \includegraphics[width=0.49\linewidth]{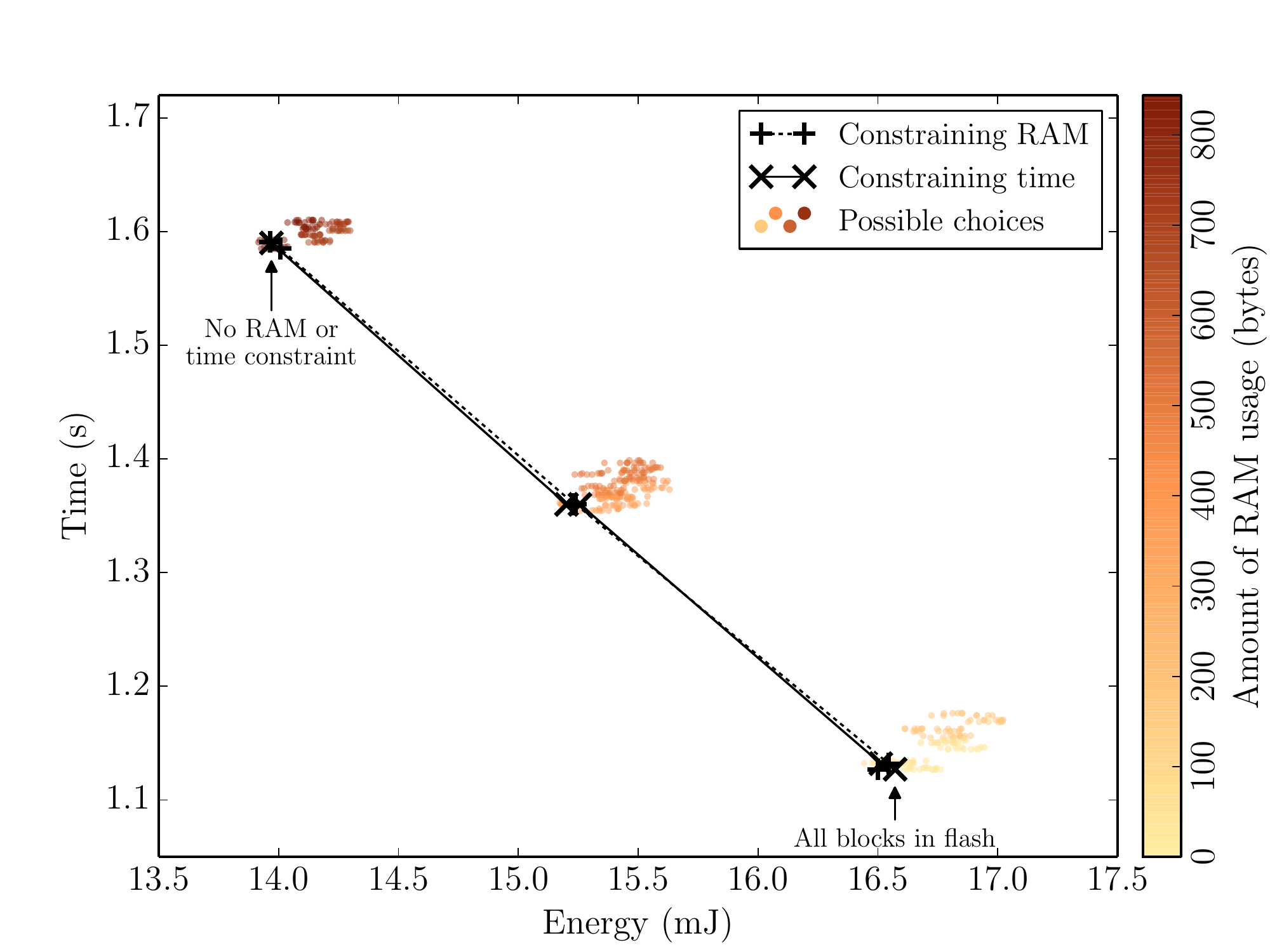}
    }
    \caption{These diagrams show the trade-off space for possible sets of basic blocks in RAM. Each point represents a possible combination of basic blocks put into RAM, along with its energy, time and RAM required. The solid line shows the solutions selected when changing the $R_{spare}$ parameter. The dashed line shows the solutions selected when changing the $X_{limit}$ parameter.}
    \label{fig:exhaustive}
\end{figure*}

The optimization was evaluated using BEEBS~\cite{Pallister2013b}, a benchmark suite designed to analyze energy consumption characteristics of the processor. The benchmark suite consists of ten programs taken from different areas of embedded applications. All of the benchmarks were measured on a STM32VLDISCOVERY board instrumented with energy measurement equipment.

The optimization was run on the set of benchmarks at the \texttt{O0}, \texttt{O1}, \texttt{O2}, \texttt{O3} and \texttt{Os} optimization levels (compiled with GCC 4.8.2). Across all benchmarks and optimization levels, the average reduction in energy and power is 7.7\% and 21.9\% respectively. The execution time is increased by an average of 19.5\%. This indicates that the optimization is effective for a wide range of benchmarks across different combinations of optimizations.

The result of applying the optimization to this benchmark suite at the \texttt{O2} and \texttt{Os} optimization levels are shown in Figure~\ref{fig:results}. This graph shows the percentage change in execution time and energy consumption when compared to the program without the optimization pass applied. Also shown are the results when an actual basic block frequency is used, as opposed to an estimate. Overall the optimization manages to decrease the energy significantly in many cases, with up to 22\% reduction in energy consumption in some cases (\texttt{int\_matmult, O2}). It is notable that this is successful despite the additional instructions executed and the overall increase in execution time.

In all cases, the average power is reduced. In some cases, the reduction is significant: 41\% reduction in the case of \texttt{fdct} at \texttt{O2} optimization level. This reduction in power is large due to both the increase in execution time and the decrease in energy consumption. As such, it occurs even when the energy is not reduced significantly. Applications which must not exceed a certain peak power will find this beneficial.

Some of the benchmarks show very little improvement (\texttt{cubic}, \texttt{float\_matmult}). These benchmarks make heavy use of library calls and emulated floating point calculations. The library calls are statically linked into the final executable and the optimization pass does not see these functions, so cannot place them into RAM. This limitation could be removed if the optimization pass was moved into the linker, allowing it to operate on all emitted code.

In all of the cases, the results are very similar when the basic block frequency is estimated, versus the actual frequencies. This demonstrates that a static estimate is good enough to achieve good results, without having to go through the lengthy procedure of instrumenting the application for profiling, or simulating it.

The choice of basic blocks to go into RAM made by the ILP solver is not necessarily the optimal solution. However, it is usually a good solution, out of the possible choices. Figure~\ref{fig:exhaustive} shows the space of possible solutions ($2^k$ solutions, where $k$ is the number of basic blocks). This space shows the energy usage and execution time of each solution, along with the RAM usage of each solution (colored). The point marked `All blocks in flash' is the base case, where no basic blocks have been moved to RAM. The point marked `No RAM or time constraint' is the solution chosen when no constraints are placed on the RAM usage, or the execution time overhead.

The dashed line shows the choices made by the solver as the RAM constraint is relaxed. For the \texttt{int\_matmult} benchmark (Figure~\ref{fig:exhaustive_int_matmult}), the solver identifies good solutions to reduce the energy consumption, while avoiding clusters of low energy but much higher execution time. Similarly, as the solver is allowed to increase the overhead (thus increasing execution time), solutions with lower energy consumption are found. This is shown by the solid line. This graph has several clusters because the benchmark has 3 basic blocks with a large size and iteration count. There are $2^3=8$ combinations of these 3 basic blocks in RAM or flash, and each of these combinations forms a separate cluster in the graph (the larger bottom cluster is formed of two combinations).

The graph for the \texttt{fdct} benchmark (Figure~\ref{fig:exhaustive_fdct}) shows similar effects, with more energy efficient, but slower solutions being found as the constraints are relaxed. This graph only has three clusters, because there are two large and similarly sized basic blocks. When neither block is in RAM, the points are clustered in the bottom right of the graph. When both are in RAM, then the points are in the cluster at the top left. Otherwise, when just one of the blocks is selected the solutions are in the middle.

\section{Case study}
\label{sec:case_study}

\begin{figure}
    \subfloat[The power profile of an application which periodically wakes to perform computation.\label{fig:casestudy_initial}]{
        \includegraphics[width=0.99\linewidth]{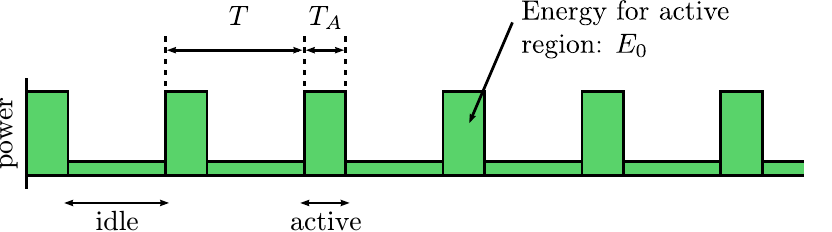}
    }

    \subfloat[The power profile of the application after the optimization is applied.\label{fig:casestudy_opt}]{
        \includegraphics[width=0.99\linewidth]{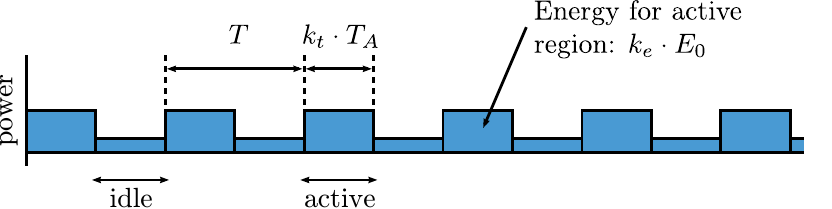}
    }
    \caption{Power profiles for an application before and after the optimization is applied. The application periodically executes some code (the active region) then waits in an sleep state until the end of the period.}
\end{figure}

The optimization is particularly useful for some types of applications that these deeply embedded processors are typically used for, such as periodic sensing. In this application, the processor spends the majority of its time in a sleep mode, and wakes infrequently to perform some processing.

In this section we consider the scenario where the processor must wake up every $T$ seconds, and transform as signal using a Finite Discrete Cosine Transform (FDCT). The initial power profile for this is shown in Figure~\ref{fig:casestudy_initial}.

The total energy, $E$, for one period, $T$, of the application is given below,

\begin{equation}
    E = E_0 + P_S\cdot (T - T_A),
\end{equation}
where $E_0$ is the energy consumed by the active region of code, $T_A$ is the length of time spent in the active region (for one time period, $T$) and $P_S$ is the average power of the sleep state (quiescent power).

When the optimization is applied, the energy consumption is:

\begin{equation}
    E' = k_e\cdot E_0 + P_S \cdot (T - k_t T_A),
\end{equation}
where $k_e$ and $k_t$ are factors describing how the optimization affects the energy and time respectively. As such we expect $k_e \leq 1$ and $k_t \geq 1$ most of the time. This corresponds to the optimization reducing the energy consumption while increasing the execution time, as seen in the experimental results.

From $E$ and $E'$ we can calculate the energy saved by applying the optimization to this case example. Here, $E_s$ is the energy saved:

\begin{equation}
    \begin{array}{ll}
        E_s &= E - E' \\
            &= E_0 (1 - k_e) + P_S T_A (k_t - 1). \\
    \end{array}
    \label{eq:energy_saved}
\end{equation}

\begin{figure}
    \centering
    \includegraphics[width=0.8\linewidth]{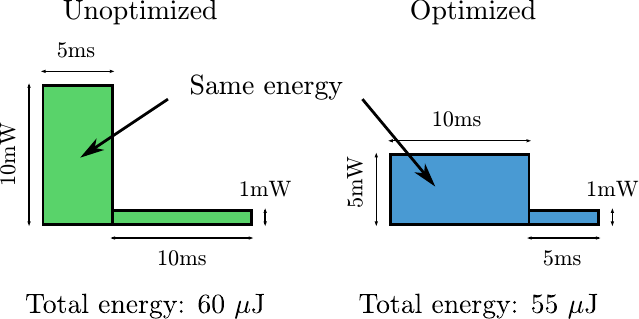}
    \caption{Diagram illustrating how keeping the energy consumption of the active region constant while the execution time increases affects the total energy. The beginning region is the active code, which is optimized.}
    \label{fig:increase_time_example}
\end{figure}

From this equation in can be seen that either decreasing $k_e$ (i.e. reducing the energy) or increasing $k_t$ (i.e. increasing the execution time) will maximize the amount of energy saved. An overall energy saving can be achieved even if the energy of the active region has not been reduced, in cases where the energy is similar and the execution time is increased. This phenomenon is shown in Figure~\ref{fig:increase_time_example}. In this diagram the active region where code is being executed requires the same amount of energy, but takes a longer time to execute in the optimized example. This leads to a reduction in the amount of time that is spent in the sleep state, and an overall decrease in energy consumption. Overall the energy is reduced from $60\mathrm{~\mu J}$ to $55\mathrm{~\mu J}$ in this illustration.

For certain benchmarks the energy saved is low (see Section~\ref{sec:evaluation}), compared to the increase in execution time. However, if this was used in a real application which required sleeping after executing the function, this will result in an overall reduction in energy consumption. The sleep power consumption for the SoC used (STM32F103RB) to prototype this optimization is measured at $P_S = \mathrm{3.5~mW}$. The values of the energy and time, as seen in Figure~\ref{fig:results} are (for \texttt{fdct}):

\begin{equation}
    \begin{array}{rcl}
        E_0 &=& 16.9 \mathrm{~mJ} \\
        T_A &=& 1.18 \mathrm{~s} \\
        k_e &=& 0.825 \\
        k_t &=& 1.33. \\
    \end{array}
\end{equation}

Substituting these values into Equation~\ref{eq:energy_saved} gives a total energy saved of $E_s = 4.32\mathrm{~mJ}$.

\begin{figure}
    \centering
    \includegraphics[width=\linewidth]{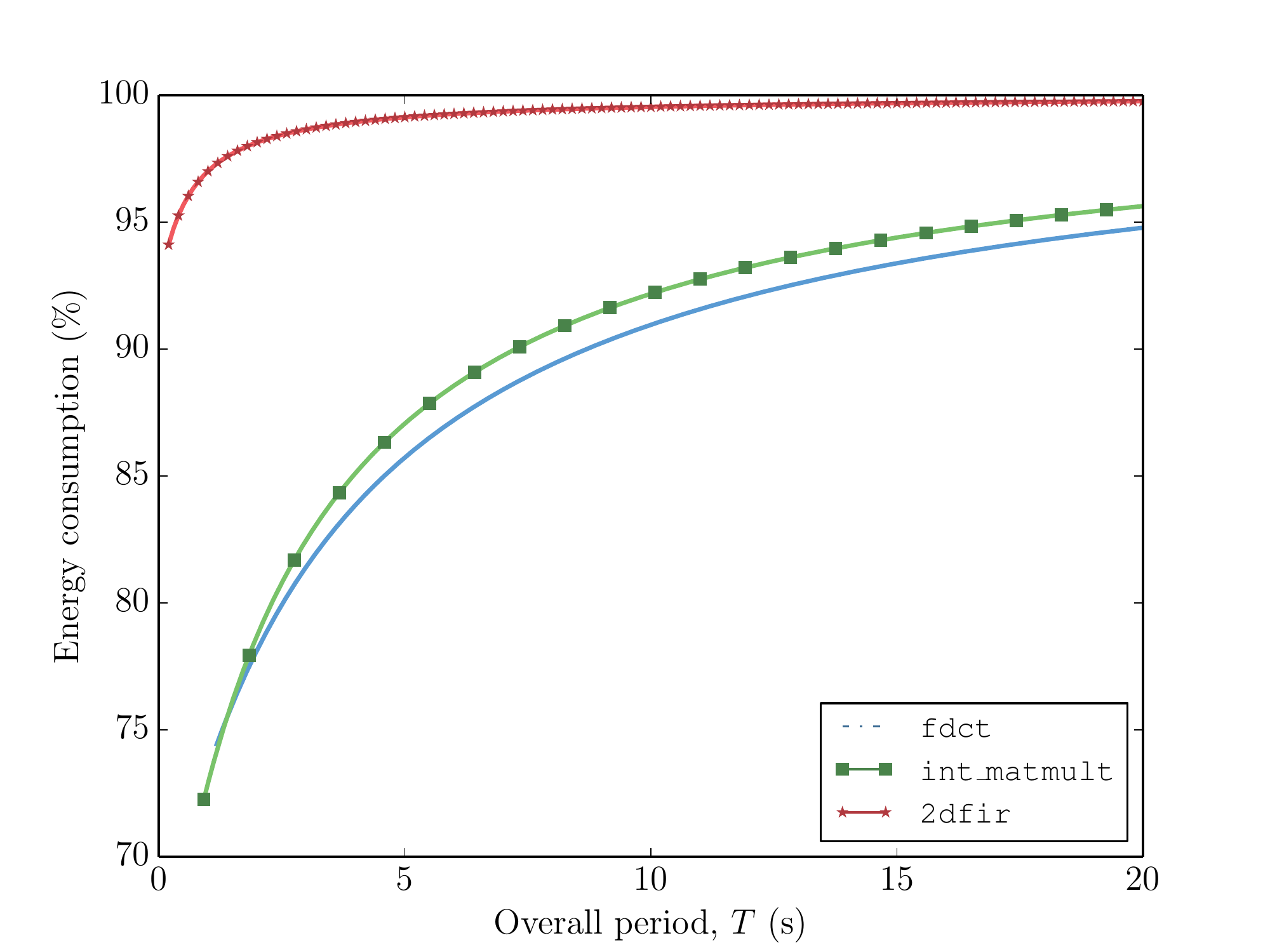}
    \caption{The proportion of energy after the optimization has been applied for different time periods, $T$. The points indicate multiples in time of the active region. I.e. the first point is for $T=T_A$ (no sleep period) and the second point is for $T=2\cdot T_A$ (with $T_A\mathrm{~s}$ of sleep time).}
    \label{fig:caseexample_savings}
\end{figure}

The energy savings as a percentage of total energy consumption cannot be calculated without knowing the value of the time period, $T$, which varies from application to application. Figure~\ref{fig:caseexample_savings} shows how $T$ affects the final energy consumption of several benchmarks. Intuitively, smaller time periods are more greatly affected by the savings achieved in the active region, since the active region is a higher overall percentage of the processor's activity.

Figure~\ref{fig:caseexample_savings} also shows other benchmarks used as the active region in this scenario. In the results graph for the optimization (Figure~\ref{fig:results}), \texttt{2dfir} did not achieve any significant energy saving but the execution time was increased. This can still result in an overall reduction in energy consumption, as seen when in this scenario, although the improvement becomes minimal as $T$ increases.

The optimization is very beneficial for this class of applications, providing up to 25\% reduction in energy consumption. This leads to up to 32\% longer battery life for devices which perform periodic sensing.

\section{Conclusion}
\label{sec:conclusion}

An optimization is presented that exploits the spare RAM often available in deeply embedded SoCs. The optimization inputs parameters into an energy cost model. This model is minimized using integer linear programming, resulting in a set of basic blocks that should be moved into RAM, to exploit its low power consumption. The code transformation is performed post-compilation, moving the required basic blocks into RAM and modifying the necessary branches to ensure the flow of execution is preserved. Significant energy savings are achieved, comparable to previous scratchpad memory studies,  despite the more restrictive low power environment.

The optimization is evaluated over all major optimization levels (using GCC), with a set of representative benchmarks. On average it reduces energy consumption by 7.7\% and power dissipation by 21.9\%. Much higher energy and power savings are seen (up to 22\% and 41\% respectively) in some cases. In particular, the savings for some of the benchmarks are limited by the current implementation of the optimization, rather than the technique itself. If given greater visibility of all the code (such as library and compiler intrinsic calls), then a greater reduction will be seen.

The techniques implemented result in a trade-off between execution time and energy consumption, since moving the code to RAM necessarily results in an overhead. The execution time increased by an average of 19.5\% on our platform. Despite this, the code relocation manages to reduce the overall energy consumption. The increase in execution time is not problematic for a large class of applications which run on deeply embedded SoCs. A case example of periodic sensing, where the processor will perform a computation then return to sleep periodically was demonstrated. The combination of a reduction in energy consumption and increase in execution time was shown to be beneficial, resulting in larger energy consumption savings (up to 25\%) than the savings achieved without the context of the application (22\%). As a result, the battery life of the device can be extended by up to 32\%.


\subsection*{Future work}

The current prototype of this optimization cannot operate on library code and other code which is used at link time. The optimization could be moved into the linker, allowing it to have a full view of the program. This should enable library code to be moved into RAM as well, improving the results since all of the basic blocks can be moved into RAM and the entire structure of the program is accounted for.

\bibliographystyle{IEEEtranS}
\bibliography{library}

\end{document}